\documentclass[
twocolumn,
amsmath,
amssymb,
floatfix,
superscriptaddress,
aps,
prl
]{revtex4-2}
\usepackage{graphicx}
\usepackage{dcolumn}
\usepackage{bm}
\usepackage{color}
\usepackage{amssymb}
\usepackage{amsmath}
\usepackage{xfrac}
\usepackage{mathrsfs}
\usepackage{comment}
\usepackage[normalem]{ulem}

\usepackage{hyperref}
\usepackage[dvipsnames]{xcolor}
\hypersetup{
    colorlinks=true,
    linkcolor=Blue,
    citecolor=Blue,   
   urlcolor=Blue
   }

\def\bea{\begin{eqnarray}}
\def\eea{\end{eqnarray}}
\def\beq{\begin{equation}}
\def\eeq{\end{equation}}
\def\f{\frac}

\def\a{\alpha}
\def\g{\gamma}

\def\rv{\mathbf{r}}
\def\nv{\mathbf{n}}
\def\Fv{\mathbf{F}}
\def\vs{\mathbf{v}_{\rm shear}}
\def\gd{\dot{\gamma}}

\begin{document}

\title{Tuning Steady Shear Rheology through Active Dopants}

\author{Amir Shee}%
\email[Contact author: ]{amir.shee@uvm.edu}
\affiliation{Department of Physics, University of Vermont, Burlington, VT 05405, USA}

\author{Ritwik Bandyopadhyay}
\affiliation{Department of Physics, University of Vermont, Burlington, VT 05405, USA}

\author{Haicen Yue}
\email[Contact author: ]{haicen.yue@uvm.edu}
\affiliation{Department of Physics, University of Vermont, Burlington, VT 05405, USA}

\begin{abstract}
We numerically investigate the steady shear rheology of mixtures of active and passive Brownian particles, with varying fractions of active components. We find that even a small fraction of active dopants triggers fluidization with comparable efficiency to fully active systems. A combined parameter, active energy, given by dopant fraction multiplied by propulsion speed squared controls the steady shear rheology and glass transition of the active-passive mixtures. These results together provide a quantitative strategy for fine-tuning the mechanical properties of a soft material with small amounts of active dopants.
\end{abstract}

\maketitle


The mechanical properties of disordered soft materials significantly change during the transition between fluid-like and solid-like states, as observed in systems ranging from synthetic colloidal suspensions \cite{Klongvessa2019}, emulsions \cite{Zhang2005}, foams \cite{Durian1995, Katgert2008, Katgert2009}, and granular matter \cite{Abate2006, Keys2007, Majmudar2007, Morse2021} to dense biological tissues\cite{Angelini2011} and embryonic development \cite{Schotz2013}. The transition in passive systems can be controlled by factors such as density, external forces, microstructural organization, interparticle interactions, and temperature \cite{Pusey1986, Liu1998, Coussot2005, Berthier2011, Liu2011, Ikeda2012PRL}. Recent studies have demonstrated that active matter, consisting of self-propelled components, exhibits unique rheological behavior distinct from passive systems, leading to novel flow properties and phase transitions \cite{ten-Hagen2011, Peshkov2016, Martens2017, Asheichyk2019, Mitchel2020, Mandal2021, Reichert2021, Madden2022, Wiese2023PRL, Bayram2023, Amiri2023}.

With detailed balance breaking, active matter has shown special behaviors and phenomena, such as motility-induced phase separation (MIPS), giant number fluctuations (GNF), hyperuniformity, non-equilibrium glass transition (active glass) \cite{Berthier2013, Ni2013, Berthier2014, Levis2015, Bi2016, Berthier2017, Janssen2019, Henkes2020, Paoluzzi2022, Keta2022}. In addition, active suspensions show dramatic rheological changes, including reduced viscosity~\cite{Wiese2023PRL, Mo2024}, shear thinning (and at high activity even shear thickening)~\cite{Bayram2023}, and shear-induced orientation ordering~\cite{Mandal2021}, and motility-induced velocity ordering~\cite{Abbaspour2024}. 
Aiming to develop a general understanding of fluid-glass-jamming transitions in amorphous materials, the role of activity, as another source of energy to overcome local energy barriers, has also been compared to temperature \cite{Wiese2023PRL} and external shear \cite{Mo2020,Peter2021,Agoritsas2021,Carlos2021}. A distinctive feature of activity, as a local drive, is that it can be manipulated at the particle level, instead of system level. This enables the introduction of activity heterogeneity into the system, with the simplest case being a binary mixture of active and passive particles. Previous work has shown that small amounts of active particles can induce motility-driven phase separation \cite{Schwarz-Linek2012} or even promote crystallization in disordered materials \cite{Ni2014}.   
However, the quantitative role of various fractions of active doping in triggering the solid-to-fluid transition remains an open question.

Here, we investigate how small-amount active doping controls the glass‑to‑fluid transition. We identify ``active energy", given by the dopant fraction multiplied by propulsion speed squared as the control parameter of transition. Remarkably, even a small dopant fraction (well below 50 percent) suffices to fluidize systems with comparable efficiency to fully active systems, revealing the non-negligible role of even a minimal amount of active components and suggesting a strategy to finetune soft materials' properties with them. 

\begin{figure}[!b]
\begin{center}
\includegraphics[width=0.75\linewidth]{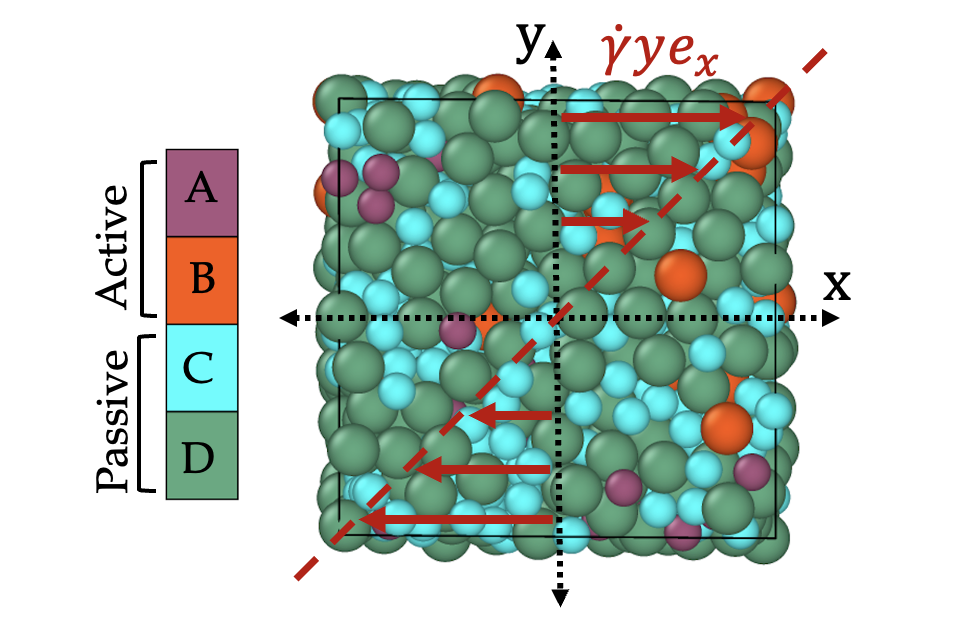} 
\caption{The shear velocity profile is shown on a cross-section in the x-y plane of the three-dimensional simulation box, representing a small fraction of active particles doped in passive particles.} 
\label{fig:schematic}
\end{center}
\end{figure}
We simulate three-dimensional binary mixtures of soft active and passive Brownian particles under different shear strain rates and systematically varying the fraction of active Brownian particles i.e., active dopants. Here, we consider a dry active matter model to decipher the role played by self-propulsion alone \cite{Wysocki2014, Shaebani2020, Turci2021, Wiese2023PRL}, excluding the fluid-mediated hydrodynamic forces and particle-wall interactions that are critical in dilute microswimmer suspensions \cite{Rafai2010, Gachelin2013, Lopez2015, Saintillan2018, Papadopoulou2020, Martinez2020, Xu2024, Burriel2024}. Moreover, hydrodynamic contributions are expected to become negligible at high densities, as observed in colloidal glasses \cite{Hunter2012} and confluent cell tissues \cite{Bi2016, Lawson-keister2021, Henkes2020} and absent in dry active systems \cite{Deseigne2010, Briand2018, Scholz2018, Baconnier2022}. 
\begin{figure*}[!t]
\begin{center}
\includegraphics[width=8.5cm]{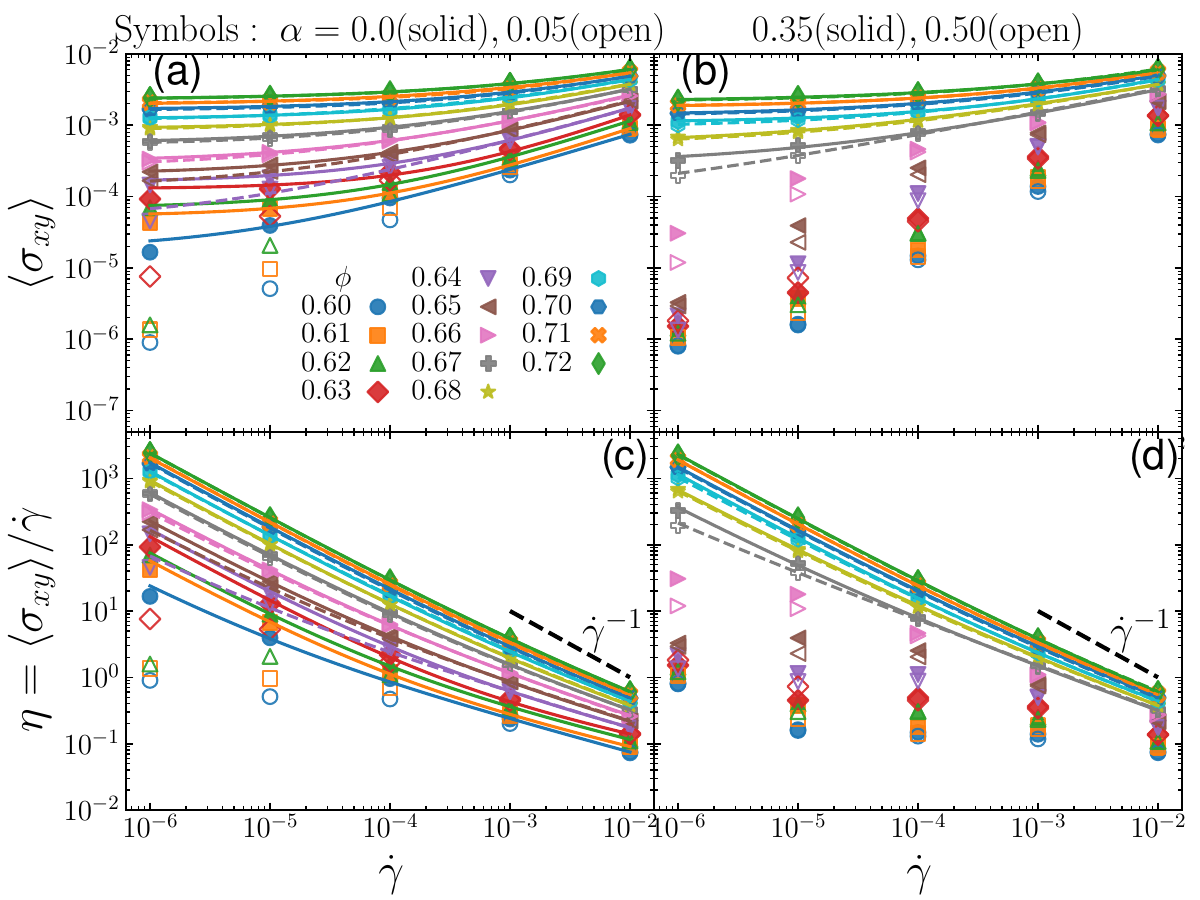} 
\includegraphics[width=8.5cm]{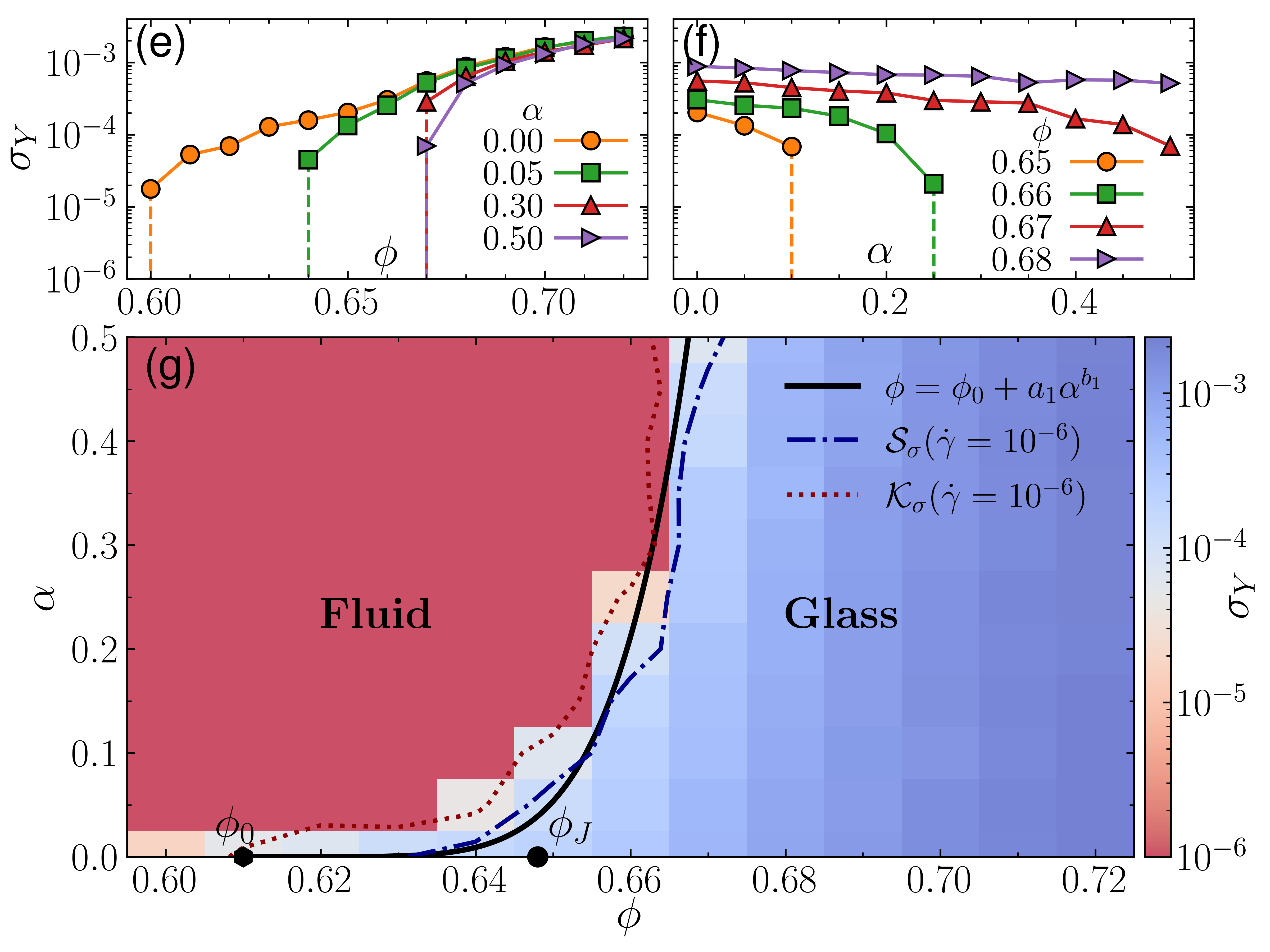} 
\caption{
Steady Shear Glass-Fluid Rheology with varying packing density($\phi$) and active fraction($\alpha$) with active strength (i.e., P\'eclet number) $\mathrm{Pe}=10$. Flow curves (a, b) the shear stress, $\langle\sigma_{xy}\rangle$, and (c, d) the viscosity, $\eta$, at densities $\phi = 0.60, \ldots, 0.72$. (a) and (c) depict the behavior for two active fractions: $\alpha = 0.0$ (purely passive; solid symbols and solid lines) and $\alpha = 0.05$ (small active fraction; open symbols and dashed lines). (b) and (d) illustrate the results for two relatively high active fractions, $\alpha = 0.35$ (solid symbols and solid lines) and $\alpha = 0.50$ (open symbols and dashed lines). The lines represent fits of the Herschel-Bulkley form in Eq.~\eqref{eq:Herschel_Bulkley_Yield}; used to extract yield stress $\sigma_{Y}$. Yield stress $\sigma_{Y}$ as function of packing density $\phi$ in (e) and active fraction $\a$ in (f). (g) Phase diagram in $\phi-\a$ plane. The black solid line corresponds to shifted power law $\phi(\a) =\phi_{0} + a_1 \a^{b_1}$ where $a_1,b_1$ are the fitting constants ($a_1\approx 0.064$, $b_1\approx 0.16$) with passive glass transition density $\phi_{0}=0.61$~\cite{Wiese2023PRL}. The blue dashed dotted (red dotted) line represent skewness (excess kurtosis) of shear stress fluctuations for $\dot{\gamma}=10^{-6}$ (see Fig.~3 in Supplemental Material~\cite{Supply2025}). The black solid circle at $\phi_{J}=0.648$ represent passive athermal ($\alpha=0,~k_{B}T=0$) Jamming transition~\cite{Wiese2023PRL}.
} 
\label{fig:shear_yield_stress_alpha_phi}
\end{center}
\end{figure*}

The system comprises $N$ particles, where a fraction, $\a=N_{a}/N$, are active particles. To suppress crystallization, each category of particles is further equally divided into two subtypes~\cite{Olsson2007PRL, Ikeda2012PRL, Wiese2023PRL}: small and large radii, resulting in four distinct particle types: A (small active), B (large active), C (small passive), and D (large passive); see Fig.~\ref{fig:schematic}. The particle radii are defined as follows: $R_0=0.5$ is the radii of small particles (A and C). The radii of large particles (B and D) are scaled relative to their small counterparts by a factor of $\sqrt{2}$, ensuring consistent volume ratios. The cubic simulation box length $L$, is determined to maintain the target volume fraction i.e., packing density ($\phi$): 
$L=(\sum_{k=1}^{4} N_k 4\pi R_k^3/3\phi)^{1/3}$, where $N_k$ is the number of particles of type $k$, and $R_k$ is their respective radius. The active particles(A,B) evolve with overdamped active Brownian particles (ABPs) dynamics
\bea
\dot{\rv}_i &=& v_0 \nv_i + \mu_0 \Fv_i + \sqrt{2 D_t} \boldsymbol{\xi}_{i}(t)+ \gd y_i \boldsymbol{e}_x\,,\label{eom1}\\
\dot{\nv}_i &=& \sqrt{2 D_r} \nv_i \times \boldsymbol{\nu}_{i}\,.
\label{eom2}
\eea
The first term, $v_0 \nv_i$, represents the self-propulsion with active speed $v_0$, which acts along the orientation unit vector $\nv_i$. For passive particles (C,D), we consider $v_0=0$. The second term, $\mu_0 \Fv_i$ is the interaction force on $i$-th particle defined as $\Fv_i=\sum_{j\neq i} \Fv_{ij}$, where $\Fv_{ij}$ is the soft repulsive interaction force between $i$-th particle with neighboring $j$-th particle. In our simulations, particle interactions are modeled as harmonic repulsions, approximated using a Morse potential (detailed in Supplemental Material~\cite{Supply2025}). The mobility coefficient $\mu_0=1$. Both $\boldsymbol{\xi}_i(t)$ and $\boldsymbol{\nu}_i(t)$ are Gaussian white noise terms with zero mean and unit variance.
The translational diffusion coefficient $D_t=\mu_0 k_B T$ with $k_{B}T=10^{-4}$, ensuring that the fluctuation-dissipation theorem is satisfied in the absence of activity and shearing.
The rotational diffusion coefficient is $D_r=3D_{t}/a_0^2$, where $a_0=2R_0$ is the diameter of small particles.  
We impose a linear velocity profile on the particles $\vs^i = (\gd y_i, 0, 0)$; see Fig.~\ref{fig:schematic} and apply Lee-Edwards boundary conditions~\cite{Lees1972, Dhont1996, Coussot2005, Allen2017}. The dynamical Eqs.~\eqref{eom1} and \eqref{eom2} are integrated using Euler-Mayurama scheme~\cite{Raible2004, Snook2006} with a time step $dt=10^{-2}$ implemented in HOOMD‑blue~\cite{Anderson2020} and also described explicitly in Supplemental Material~\cite{Supply2025}.

The key parameters to vary include volume fraction $\phi$, active fraction $\a=N_a/N$, P\'eclet number ${\rm Pe} = v_0 / a_0 D_r$, shear rate $\gd$. Throughout our study, we hold either $\phi$ or ${\rm Pe}$ constant and vary the other three parameters. We explore $\gd=10^{-6},\ldots,10^{-2}$, $\a=0.0,\ldots,0.5$, $\mathrm{Pe}=0,\ldots,20$, and $\phi=0.60,\ldots,0.72$. These parameter ranges prevent motility‐induced phase separation~\cite{Wysocki2014, Turci2021, Omar2021}. 

The effect of activity on the glass-fluid transition is quantified by systematic measurements of shear stress $\langle\sigma_{xy}\rangle$ and corresponding viscosity $\eta=\langle\sigma_{xy}\rangle/\dot{\gamma}$ under different shear rate $\dot{\gamma}$. We first keep the P\'eclet number fixed at $\mathrm{Pe}=10$ while varying the packing fraction $\phi$ and active fraction $\alpha$. For the limiting cases $\a=0$ i.e., purely passive systems, the fluid-to-solid transition, marked by the emergence of a finite yield stress, occurs at $\phi_0=0.61$ \cite{Ikeda2012PRL}, while for fully active systems ($\a=1$) with ${\rm Pe}=10$, the transition is pushed to $\phi_1=0.68$ \cite{Wiese2023PRL}. Here, we investigate the cases in between to understand how partial active components tune the phase transition between these two limits. Figures~\ref{fig:shear_yield_stress_alpha_phi}(a-d) show flow curves of shear stress, $\langle\sigma_{xy}\rangle(\dot{\gamma})$, and viscosity, $\eta(\dot{\gamma})$, respectively, for $\phi$ spanning $0.60-0.72$ at four representative dopant levels: $\alpha=0$ and $\alpha=0.05$ in (a,c) and $\alpha=0.35$ and $\alpha=0.5$ in (b,d). To pinpoint the glass-to-fluid transition, we extract the yield stress, $\sigma_{Y}$, by fitting each $\langle\sigma_{xy}\rangle(\dot{\gamma})$ curve to the Herschel–Bulkley model~\cite{Mullineux2008} 
\bea
\langle\sigma_{xy}\rangle(\dot{\gamma})=\sigma_{Y}+k \dot{\gamma}^n\,.
\label{eq:Herschel_Bulkley_Yield}
\eea
In the solid regime, a finite yield stress corresponding to the low $\dot{\gamma}$ plateau can be obtained while in the fluid regime with no obvious plateau, the fitting fails and we set $\sigma_{Y}=10^{-6}$ (negligible). These results are summarized in Fig.~\ref{fig:shear_yield_stress_alpha_phi}(e-g). We see in (e) that even a $5\%$ active fraction dramatically changes the glass transition compared to the passive limit, whereas a more substantial increase from $35\%$ to $50\%$ does not affect the transition as much. This reveals that even minimal active doping exerts a pronounced influence over the steady shear rheology of dense soft materials, and crossing some threshold fraction, the fluidization effect of active dopants almost saturates. This points to a possible strategy for enhancing efficiency and lowering costs in materials engineering through the controlled use of active dopants. Fig.~\ref{fig:shear_yield_stress_alpha_phi}(f) gives the minimal active fraction required to fluidize a system with a given packing density. 

These results can be summarized as a color map of $\sigma_Y$ on the $\phi-\alpha$ plane as in Fig.~\ref{fig:shear_yield_stress_alpha_phi}(g). We fit the crossover $\phi_G (\alpha)$ to a shifted power law $\phi_G (\a) =\phi_{0} + a_1 \a^{b_1}$ where $\phi_{0}=0.61$, the glass transition density for passive case~\cite{Wiese2023PRL} with fitting constants $a_1\approx 0.064$, $b_1\approx 0.16$ (the black solid line). The value of $a_1$ can also be directly estimated from Wiese et al.\cite{Wiese2023PRL} as $a_1=\phi_1-\phi_0\approx 0.07$ with transition density for fully passive ($\phi_0$) and fully active cases ($\phi_1$), closely matching our estimated value. The value of $b_1$ indicates a strong nonlinear dependence of the glass transition density $\phi$ on effective active energy $\alpha v_0^2$, in contrast to the glass transition temperature, which exhibits a linearly dependence on $v_0^2$~\cite{nandi2018random}. We also show that the probability distribution shape of the shear stress, which can be quantified using high order cumulants like the skewness $\mathcal{S}_{\sigma}$ and excess kurtosis $\mathcal{K}_{\sigma}$, are sensitive to the glass-fluid transition, indicating a deviation from half normal distribution (Fig.2 and 3 in \cite{Supply2025}).
Both $\mathcal{S}_{\sigma}$ and $\mathcal{K}_{\sigma}$ undergo a sharper change than the average value $\langle \sigma_{xy}\rangle$ across the glass transition, as shown in Fig.~\ref{fig:stress_skew_kurt_alpha_Pe} and thus can be used as additional independent, and probably better, criteria to determine the glass-fluid transition. The blue dash–dotted and red dotted lines indicate, respectively, the fluid–glass boundaries determined from the skewness $\mathcal{S}_{\sigma}$ and the excess kurtosis $\mathcal{K}_{\sigma}$ of the shear‐stress fluctuations at $\dot{\gamma}=10^{-6}$ (see Supplemental Material~\cite{Supply2025} for detailed calculation and interpretation). 

\begin{figure}[!t]
\begin{center}
\includegraphics[width=8.5cm]{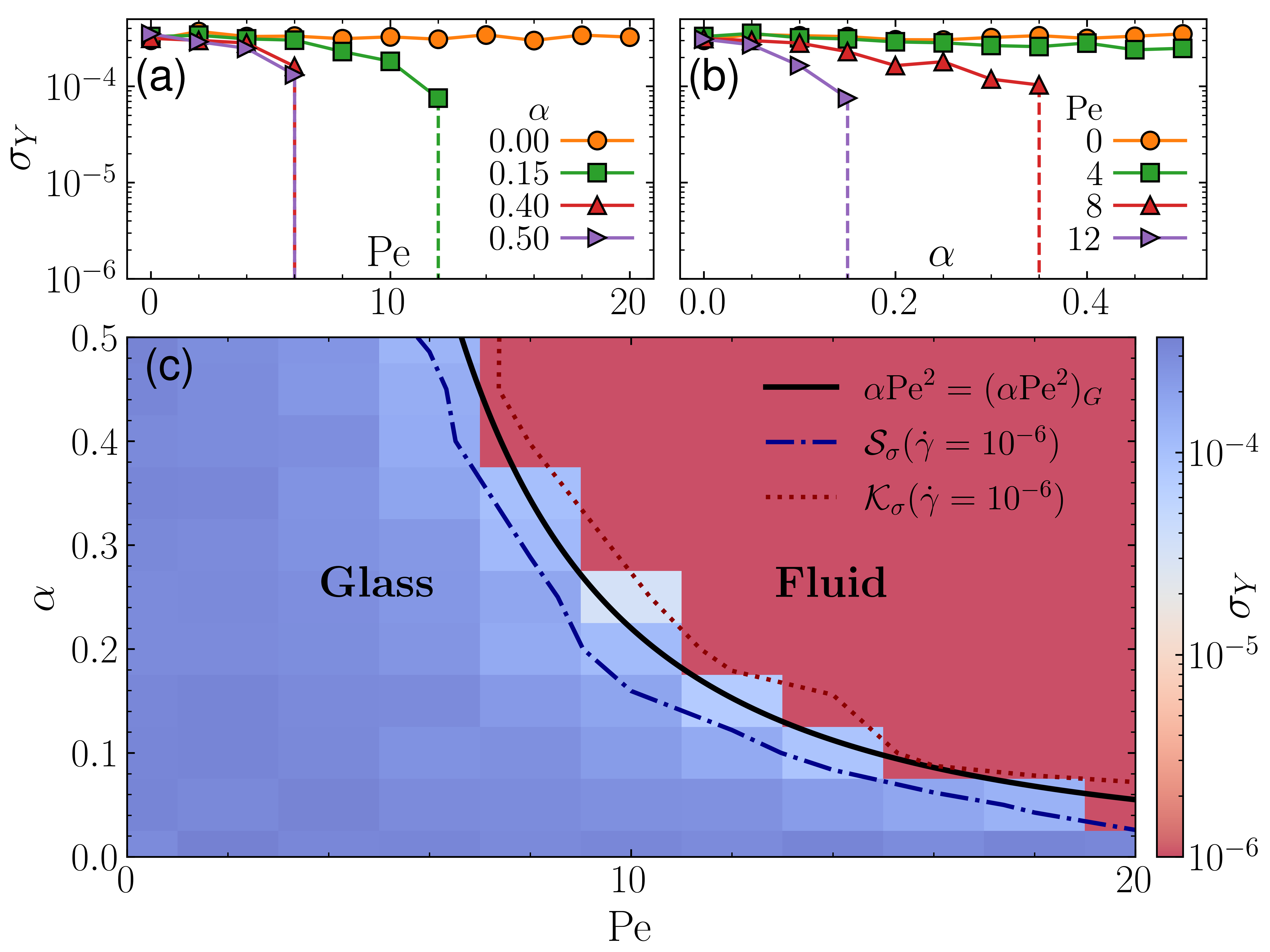} 
\caption{Yield stress with varying active strength P\'eclet $\mathrm{Pe}$ and fraction ($\a$) at density $\phi=0.66$. 
Yield stress $\sigma_{Y}$ as function of $\mathrm{Pe}$ in (a) and active fraction $\a$ in (b). Vertical dashed lines in (a,b) indicate threshold of glass transition.
(c) Phase diagram in $\mathrm{Pe}-\a$ plane. The black solid line corresponds to $\a \mathrm{Pe}^2=(\a \mathrm{Pe}^2)_G\approx 22$ where $(\a \mathrm{Pe}^2)_G$ is the critical glass transition activity. The blue dashed dotted (red dotted) line represent the glass-fluid transition calculated from skewness (excess kurtosis) of shear stress fluctuations for $\dot{\gamma}=10^{-6}$ (see Fig.~\ref{fig:stress_skew_kurt_alpha_Pe}).
} 
\label{fig:yield_stress_alpha_Pe}
\end{center}
\end{figure}

\begin{figure}[!t]
\begin{center}
\includegraphics[width=\linewidth]{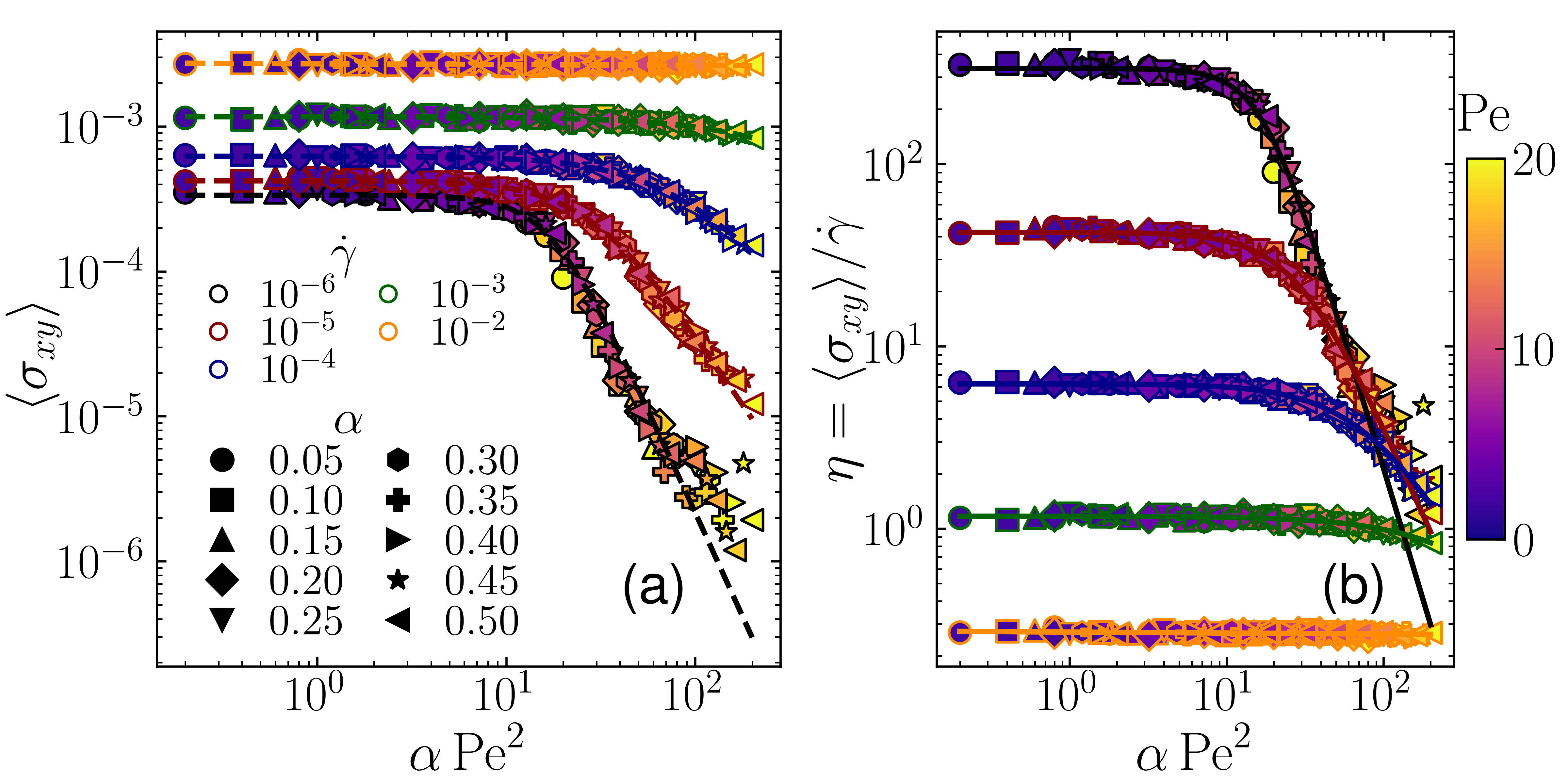} 
\includegraphics[width=\linewidth]{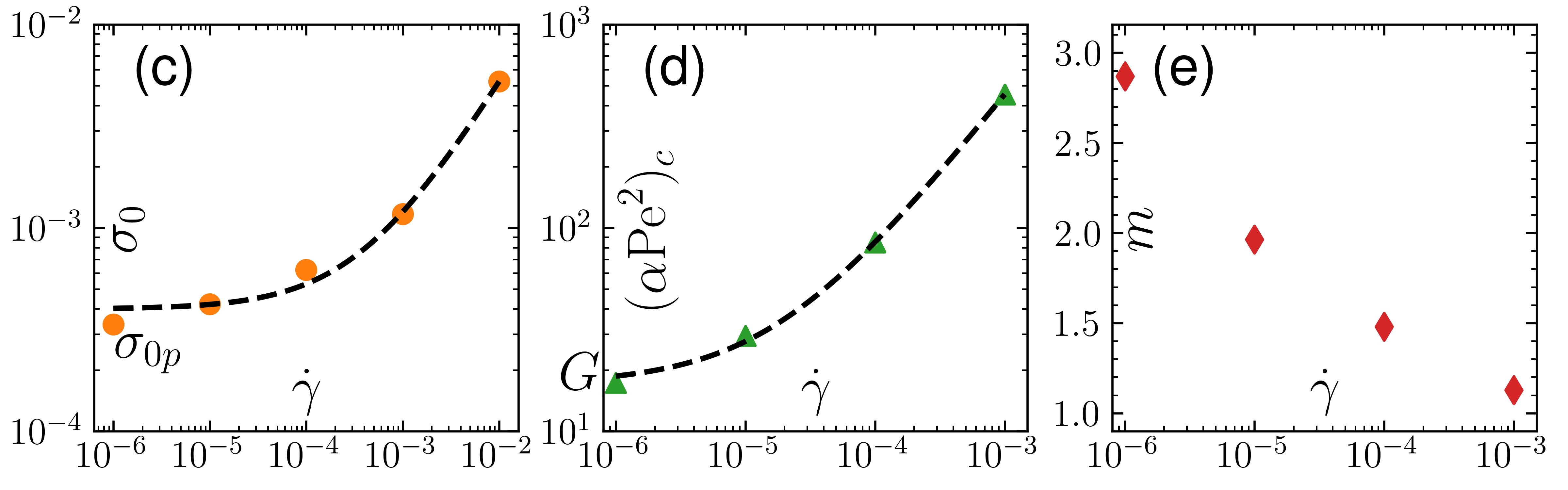}
\caption{
Impact of activity on the Glass-Fluid transition. (a) Shear stress, $\langle\sigma_{xy}\rangle$, and (b) the viscosity, $\eta$, as a function of activity ($\alpha \mathrm{Pe}^2$) at density $\phi = 0.66$ for varying shear rates $\dot{\gamma}$. The dashed lines in (a) are fits to Eq.~\eqref{eq:Hill_emperical_form}.
(c) $\sigma_0$, (d) $(\a \mathrm{Pe}^2)_c$, and (e) $m$ as a function of $\dot{\gamma}$. Symbols are values extracted from the fits in (a,b), and dashed lines are Herschel–Bulkley fits. (c) The fitted yield stress in the passive limit($\a\mathrm{Pe}^2\to 0$) is $\sigma_{0p}\approx 3.98\times 10^{-4}$. (e) show that $m$ diverge as $\dot{\g}\to 0$.
(c) stress (shear and yield) through the glass-fluid crossover. The symbols shape indicates active fraction ($\a$), color represents P\'eclet $\mathrm{Pe}$, and border color denotes shear rates ($\dot{\gamma}$).
} 
\label{fig:shear_stress_alpha_Pe}
\end{center}
\end{figure}

We then keep the packing density fixed at $\phi=0.66$ while varying both the P\'eclet number, $\mathrm{Pe}$, and the active fraction, $\alpha$. 
In Fig.~\ref{fig:yield_stress_alpha_Pe}(a), we plot yield stress $\sigma_{Y}$ versus Pe for several values of $\alpha$, while (b) shows $\sigma_{Y}$ versus $\alpha$ for several values of Pe. The curves for $\alpha=0$ in (a) and $\mathrm{Pe}=0$ in (b) serve as the purely passive reference. As either $\alpha$ (in a) or $\mathrm{Pe}$ (in b) increases, the transition shifts to progressively lower values, indicating that increasing activity in either way, by increasing activity strength or active particle fraction, can fluidize the system.
In (c), we combine these results into a color map of $\sigma_{Y}$ on the Pe–$\alpha$ plane, clearly delineating the glass ($\sigma_{Y}>0$) and fluid ($\sigma_{Y}\approx0$) regimes. In fully active systems, it is reasonable to treat activity as an effective temperature $T_{\mathrm{eff}}\propto v_0^2$ as in \cite{nandi2018random}. However, in heterogeneous systems like the active-passive mixtures, especially when the active fraction is low, it is not guaranteed that the \emph{local} activity can be represented by a \emph{global} effect temperature. Surprisingly, we find that the transition can be well captured by the relation $\a \mathrm{Pe}^2=(\a \mathrm{Pe}^2)_G$ where $\alpha\mathrm{Pe}^2$ can be understood as a dimensionless parameter characterizing the active energy $N_a v_0^2$. The black solid line represents a fitting with $(\a \mathrm{Pe}^2)_G = 22$.

In fact, the combined parameter $\alpha\mathrm{Pe}^2$ controls not only the glass transition, but also the steady-state shear rheology across a broad range around the glass transition. In Fig.~\ref{fig:shear_stress_alpha_Pe}(a,b), we collapse almost all the data above with different $\alpha$ and $\mathrm{Pe}$ onto a single master curve $\langle\sigma_{xy}\rangle(\alpha \mathrm{Pe}^2)$ or $\eta(\alpha \mathrm{Pe}^2)$, for each shear rate $\dot{\gamma}$, as shown in Fig.~\ref{fig:shear_stress_alpha_Pe}(a,b). We note that this data collapse slightly fails at the right end of the $\dot\gamma=10^{-6}$ curve. This suggests a breakdown of mean-field approximation -- where local activity is represented as a global effective temperature -- in heterogeneous systems. A more detailed discussion is provided at the end of this paper. 
Quantitative fits to Hill-type empirical form:
\bea
\langle\sigma_{xy}\rangle(\alpha \mathrm{Pe}^2) &=& \f{\sigma_{0}}{[1+(\alpha \mathrm{Pe}^2/(\alpha \mathrm{Pe}^2)_c)^m]}\,,
\label{eq:Hill_emperical_form}
\eea
capture the sigmoidal shape of the curves. In the viscosity curves (Fig.~\ref{fig:shear_stress_alpha_Pe}(b)), as $\dot{\gamma}$ decreases, the plateau becomes shorter with rising height, reflected by the increasing $\eta_0\equiv\sigma_0/\dot{\gamma}$ and decreasing $(\alpha \mathrm{Pe}^2)_c$ shown in Fig.~\ref{fig:shear_stress_alpha_Pe}(c,d). This trend suggests a diverging viscosity at the zero-shear glass transition as $\dot{\gamma}\rightarrow0$, with the increasing $m$, which characterizes the sharpness of the divergence, linked to the Mode Coupling Theory (MCT) predicted power law $\tau_\alpha\sim (v_0^2)^{-m}$ \cite{nandi2017nonequilibrium}. In the limit of high $\dot{\gamma}$ and high $\alpha \mathrm{Pe}^2$ (fluid regime), this gives an activity-induced thinning effect where viscosity scales as $\eta\sim(\alpha \mathrm{Pe}^2)^{-m}$ with $m\rightarrow1$. This can be compared to the shear-thinning scaling laws observed in many different soft materials with $\eta\sim\dot{\gamma}^{-m}$ and $m$ ranging from $0.5$ to $1$~\cite{berthier2002nonequilibrium}. 
In (c) and (d) $\sigma_0$ and $(\alpha \mathrm{Pe}^2)_c$'s dependence on $\dot\gamma$ can be fitted into the following formats:   
\bea
\sigma_{0}(\dot{\gamma}) &=& \sigma_{0p}+k_0 \dot{\gamma}^n\label{eq:fitting_passive_yield}\,,\\
(\a \mathrm{Pe}^2)_c(\dot{\gamma}) &=& (\a \mathrm{Pe}^2)_G+k_c \dot{\gamma}^n\,.
\label{eq:fitting_glass_transition}
\eea
The fitted value, $\sigma_{0p}\approx 3.98\times 10^{-4}$ gives the passive yield stress consistent with the values reported in \cite{Ikeda2012PRL,Wiese2023PRL}, while $(\a \mathrm{Pe}^2)_G\approx 17$ is the glass-to-fluid transition, roughly matching the fitted value $(\a \mathrm{Pe}^2)_G\approx 22$ in Fig.~\ref{fig:yield_stress_alpha_Pe}. 
Together, these results underscore the role of active energy $\a \mathrm{Pe}^2$ in the steady shear rheology that marks the glass-to-fluid boundary of active-passive mixtures, extending the concept of activity strength as a control parameter for the glass-to-fluid transition in purely active systems.

In summary, we have demonstrated that the glass transition density $\phi_g$ in active-passive mixtures exhibits a highly nonlinear dependence on the active fraction, with a more pronounced effect at low active fractions. This underscores the substantial impact that even a small number of active components can have on the mechanical properties of mixed soft materials, such as soils containing a small population of living organisms, or extracellular matrices embedded with a few living cells. This finding also suggests a cost-effective strategy for fine-tuning the mechanical response of soft materials through minimal active doping. We also show that the steady-state shear rheology of active-passive mixtures in a wide range across the glass transition is governed by a combined parameter $\alpha\mathrm{Pe}^2$, which quantifies the total active energy input. This observation is theoretically non-trivial considering the heterogeneous nature of the mixture. While it is common in homogeneous active systems to treat $v_0^2$ as a global effective kinetic energy (i.e., $T_{\mathrm{eff}}^2$)\cite{nandi2018random}, such a ``mean-field" approach may fail when active particles exert influence only \emph{locally}. Indeed, at the right end of the low shear rate cases (Fig.~\ref{fig:shear_stress_alpha_Pe} here, and Fig.5 in Supplemental Material~\cite{Supply2025}), the data collapse breaks down. The successful collapse of active energy across a wide range near the glass transition may be attributed to the growing correlation length scale in supercooled liquids and glasses, which allows sparse active particles to influence a larger surrounding region. Future work should aim to quantify and elucidate the conditions under which this data collapse occurs -- in other words, to determine the regime of validity for the mean-field approximation.  This combined parameter also carries practical significance. A systematic scanning in the parameter space of at least four dimensions ($\phi$,$\alpha$,$\mathrm{Pe}$,$\dot{\gamma}$) is computationally expensive. By collapsing the influence of $\alpha$ and $\mathrm{Pe}$ into a single combined parameter, the dimensionality of the problem is reduced. This simplification allows for more efficient exploration and prediction of the steady shear rheology of active-passive mixtures, and more flexibility in choosing combinations of $\alpha$ and $\mathrm{Pe}$ to suit specific experimental or application-driven constraints. 
For example, it is reported in \cite{Wiese2023PRL} that a fully active system with $\mathrm{Pe}=3$ at $\phi=0.66$ has a yield stress of $\sigma_Y\approx 3.49\times 10^{-4}$. According to our findings, this corresponds to an active energy $\alpha\mathrm{Pe}^2=9$, and should be equivalent to any mixed system with different combinations of $\alpha$ and $\mathrm{Pe}$ that yield the same active energy. This can be cross-validated using our fitted empirical form (Eq.~\eqref{eq:Hill_emperical_form}) for the $\phi=0.66$ case $\sigma_Y(9) = \sigma_{0p}/\left[1+\left(9/(\a \mathrm{Pe}^2)_G\right)^3\right] \approx 3.47\times 10^{-4}$, which closely matches the reported value. 

As the first systematic study on the steady shear rheology of the glass–fluid
boundary of active–passive mixtures, this work remains idealized, relying on simplified assumptions such as frictionless and spherical particles, negligible hydrodynamic interactions, equal sizes for active and passive particles, and isotropically directed active dopants. Future applications in specific systems will require more nuanced models that relax some of these assumptions to better capture system-specific complexities. In particular, the non-uniform distribution of active dopants, accounting for the complex structures or organization of active components, is of special interest. Theoretically, the potential interplay between such spatial heterogeneity and the intrinsic dynamical heterogeneity of glassy systems presents a compelling direction for future investigation. Additionally, size disparities between active and passive particles, which affect static structure and contact geometry, also raise important theoretical questions.

\medskip

\noindent
{\bf Acknowledgements-~}
We gratefully acknowledge the use of the Vermont Advanced Computing Cluster (VACC) at the University of Vermont for the computational resources used in this work.

\medskip

\noindent
\textbf{Data Availability Statement-~}
The simulation data supporting the findings of this study are openly available in Zenodo at Ref.~\cite{Data2025}.

\bibliography{reference} 

\section{End Matter}

\begin{figure}[!t]
\begin{center}
\includegraphics[width=\linewidth]{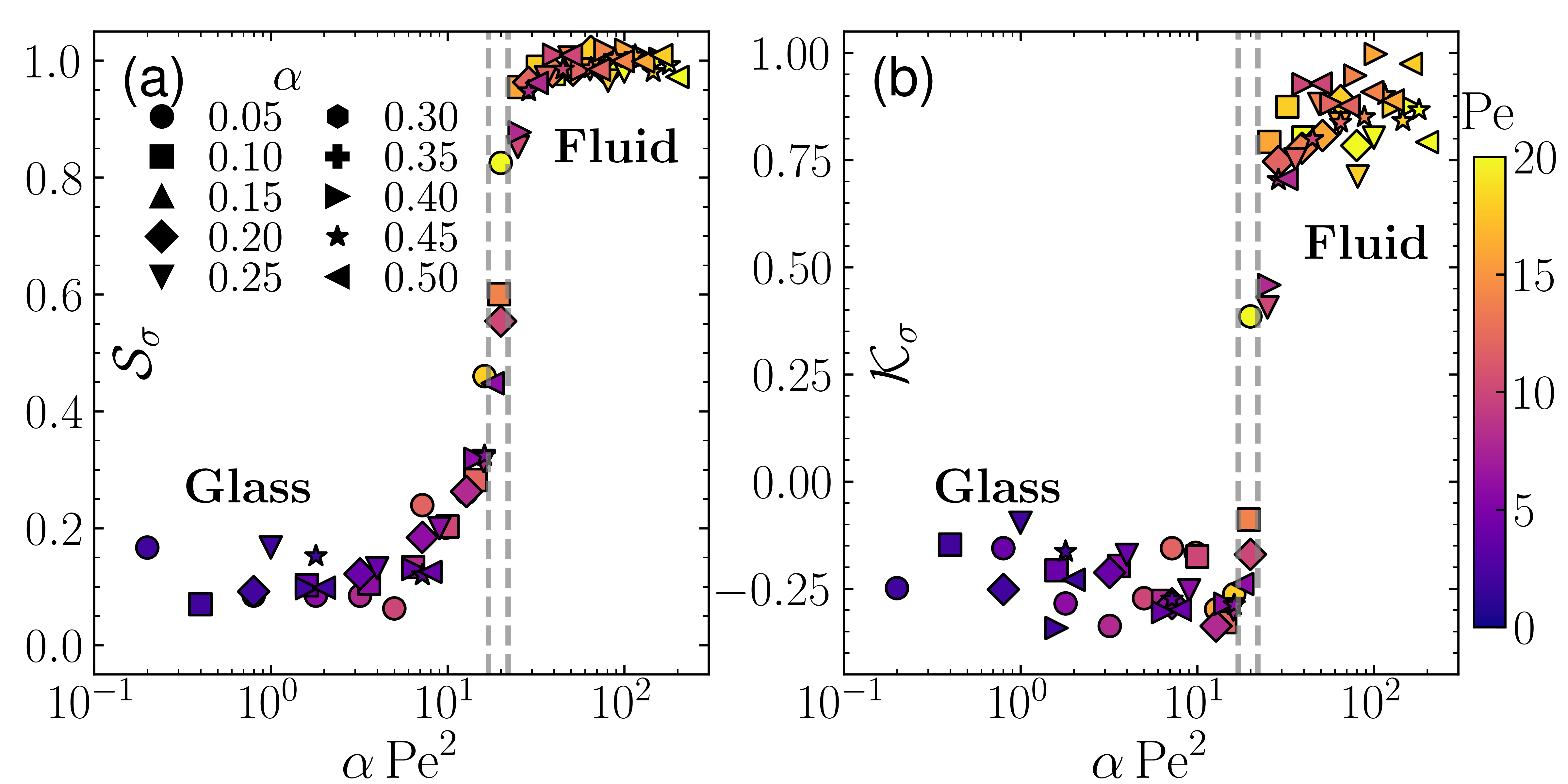} 
\includegraphics[width=\linewidth]{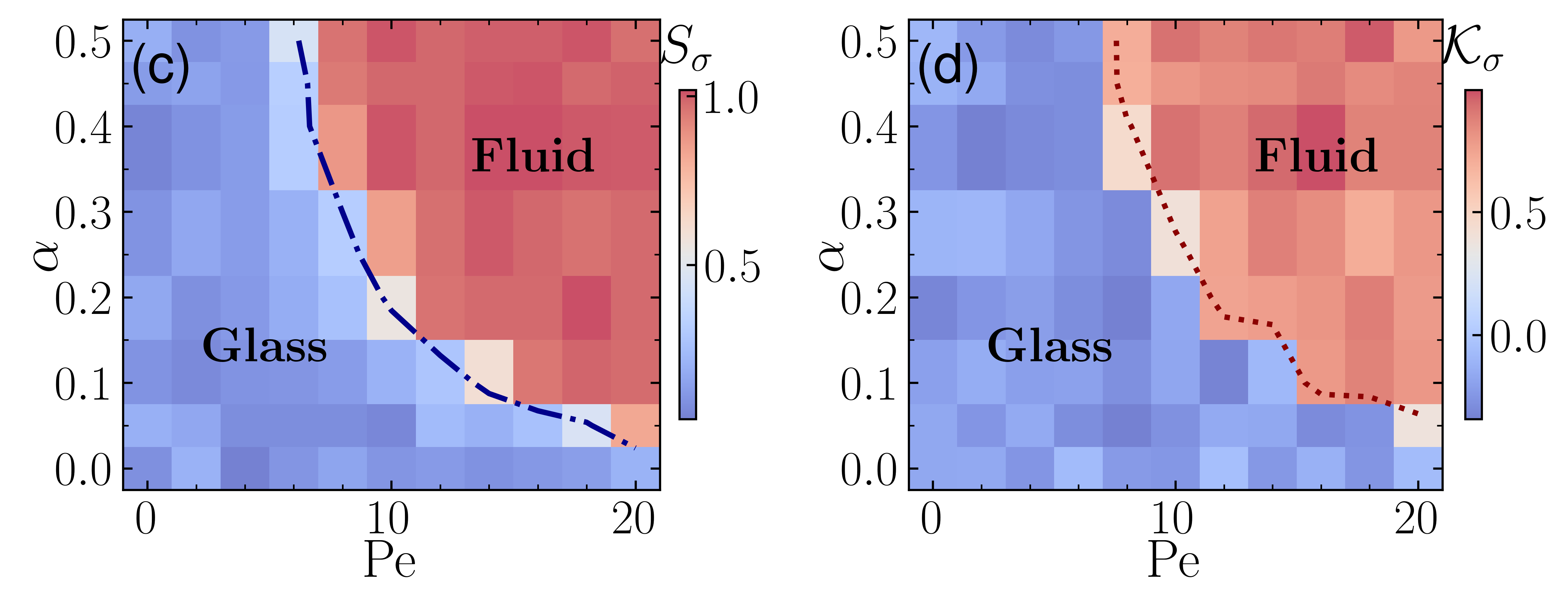}
\caption{Impact of activity on shear stress fluctuation moments. (a) Skewness $\mathcal{S}_\sigma$ and (b) excess kurtosis $\mathcal{K}_\sigma$ versus activity $\alpha\mathrm{Pe}^2$, showing the glass-fluid crossover. (c,d) Heat maps of $\mathcal{S}_\sigma$ and $\mathcal{K}_\sigma$ over the $(\mathrm{Pe},\alpha)$ plane, with $\mathcal{S}_\sigma,\mathcal{K}_\sigma=0.5$ threshold contour delineating the glass transition. We fixed $\phi=0.66$, $\dot\gamma=10^{-6}$.
}
\label{fig:stress_skew_kurt_alpha_Pe}
\end{center}
\end{figure}
\subsection{Skewness and Excess Kurtosis of Shear Stress Fluctuations}
To characterize the shape of the shear stress fluctuations $\{\sigma_{xy}(t_i)\}_{i=1}^{n_t}$, we define \emph{magnitudes} of the instantaneous shear stress
\(\sigma_{xy}(t_i)\to|\sigma_{xy}(t_i)|\) and discard the first half of the trajectory to capture long-time stationary behavior and avoiding any transient fluctuations.
For each shear rate $\dot\gamma$ and timestep $d t$, we set total timesteps to $[10/\dot\gamma]/dt,$
and discard the initial $[5/\dot\gamma]/dt$ timesteps.
Over the remaining trajectory we compute the mean shear stress, viscosity, and yield stress, as well as the skewness and excess kurtosis of the shear stress distributions.
The shear stress distributions are shown in Supplemental Material~\cite{Supply2025}. The mean and variance in shear stress
\bea
\langle \sigma_{xy}\rangle &=& \frac{1}{n_t} \sum_{i=1}^{n_t} \sigma_{xy} (t_i)~,\\~\Delta_{\sigma} &=& \frac{1}{n_t} \sum_{i=1}^{n_t} [\sigma_{xy} (t_i)-\langle\sigma_{xy} \rangle]^2\,.
\eea
The skewness $\mathcal{S}_{\sigma}$ is defined as
\bea
\mathcal{S}_{\sigma} &=& \frac{1}{n_t} \sum_{i=1}^{n_t} \left[\frac{\sigma_{xy} (t_i)-\langle\sigma_{xy}\rangle}{\sqrt{\Delta_{\sigma}}} \right]^3\,.
\eea
$\mathcal{S}_{\sigma}=0$ corresponds to a symmetric distribution of stress magnitudes around the mean, with balanced tails, $\mathcal{S}_{\sigma}>0:$ long right tail (occasional large stress magnitudes well above the mean), $\mathcal{S}_{\sigma}<0:$ long left tail (frequent small stress magnitudes below the mean).
The excess kurtosis in three dimensions defined as
\bea
\mathcal{K}_{\sigma} &=& \frac{1}{n_t} \sum_{i=1}^{n_t} \left[\frac{\sigma_{xy} (t_i)-\langle\sigma_{xy}\rangle}{\sqrt{\Delta_{\sigma}}} \right]^4 -3\,.
\eea
$\mathcal{K}_{\sigma}=0$ denotes Gaussian fluctuations; $\mathcal{K}_{\sigma}>0$ (leptokurtic): heavy tails with rare, extreme stress excursions; $\mathcal{K}_{\sigma}<0$ (platykurtic): lighter tails than Gaussian.

In the fluid regime where stress arises from a superposition of many uncorrelated collisions, the distribution of $\left|\sigma_{xy}\right|$ is effectively half normal, leading to $\mathcal{S}_{\sigma}\approx 0.995,~\mathcal{K}_{\sigma}\approx 0.869$ whenever the system behaves like a freely flowing suspension~\cite{Supply2025}. 

In the glass regime, stress distribution peaks at finite yield stress with fluctuations resulting from thermally induced or activity-driven hops, all confined by the amorphous energy landscape. Consequently: 1) Symmetric fluctuations $(\mathcal{S}_{\sigma}\approx 0)$: These hopping events produce stress increments and decrements with nearly equal probability, yielding a symmetric distribution. 2) Suppressed extremes ($\mathcal{K}_{\sigma}<0$): Finite energy barriers and elastic constraints limit large‐amplitude stress excursions. This truncation of the tails makes the distribution platykurtic (lighter tailed than Gaussian), i.e.\ excess kurtosis below zero.

In Fig.~\ref{fig:stress_skew_kurt_alpha_Pe}, we plot (a) skewness $\mathcal{S}_{\sigma}$ and (b) excess kurtosis $\mathcal{K}_{\sigma}$, which exhibit data collapse over the rescaled activity $\alpha\mathrm{Pe}^2$ for different combinations of $\alpha$ and $\mathrm{Pe}$. The plots also clearly reveal the transition from the glassy to the fluid state; the vertical dashed lines are two independent estimates of $(\alpha\mathrm{Pe}^2)_G$ from Fig.~\ref{fig:yield_stress_alpha_Pe} and Fig.~\ref{fig:shear_stress_alpha_Pe}. In (c) and (d), we show heat maps of (c) $\mathcal{S}_{\sigma}$ and (d) $\mathcal{K}_{\sigma}$, respectively, with the transition line $\mathcal{S}_{\sigma}, \mathcal{K}_{\sigma} = 0.5$ indicated (see Fig.~\ref{fig:yield_stress_alpha_Pe}(c)).

\end{document}